\documentclass[twoside,a4paper,11pt]{torus2012}
\usepackage{graphicx}
\usepackage{hyperref}
\usepackage{amsmath}
\newcommand{\micron}{\mathrm{\mu m}}
\newcommand{\la}{~^<\!\!\!\!_\sim~}
\begin{document}
\pagenumbering{arabic}
\pagestyle{myheadings}
\thispagestyle{empty}
{\flushleft\includegraphics[width=\textwidth,bb=58 650 590 680]{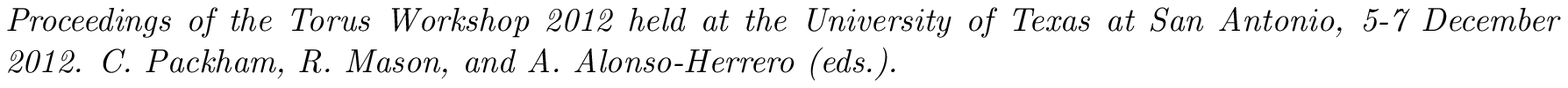}}
\vspace*{0.2cm}
\begin{flushleft}
{\bf {\LARGE
%
%%% TITLE of the paper. 
%%% TITLE of the paper. 
On donuts and crumbs: \\ 
A brief history of torus models\footnote{Slides accompanying this review can be downloaded at \url{http://tinyurl.com/donutsandcrumbs}}
%
% Do not delete next few lines
}\\
\vspace*{1cm}
%
%%% Include here the LIST OF AUTHORS.
%%% Include here the LIST OF AUTHORS.
%%% Note that the last author has to be preceeded by an AND.
Sebastian F. H{\"o}nig$^{1,2}$
%
% Do not delete next few lines
}\\
\vspace*{0.5cm}
%
%%% AFFILIATIONS LIST.
%%% and the AFFILIATIONS LIST. Note that one affiliation per line.
%%% Add as many affiliations as necessary. 
$^{1}$
Department of Physics, University of California in Santa Barbara, Broida Hall, Santa Barbara, CA 93106-9530, USA; shoenig@physics.ucsb.edu \\
$^{2}$
DFG fellow\\
%
% Do not delete next few lines
\end{flushleft}
%
% Headings
\markboth{
%%% Type the SHORT version of the paper title.
%%% Type the SHORT version of the paper title.
On donuts and crumbs
}{ % Do not delete
%
%%%  First Author \& Second Author   OR   First-author et al. 
%%%  First Author \& Second Author   OR   First-author et al. if the author list 
%%% contains three or more authors.
S.~F. H{\"o}nig
% 
% Do not delete next few lines
}
\thispagestyle{empty}
\vspace*{0.4cm}
\begin{minipage}[l]{0.09\textwidth}
\ 
\end{minipage}
\begin{minipage}[r]{0.9\textwidth}
\vspace{1cm}
\section*{Abstract}{\small
%
% ABSTRACT ABSTRACT ABSTRACT
% ABSTRACT ABSTRACT ABSTRACT
%%% Type the ABSTRACT of your paper
A variety of torus models for the infrared emission of AGN has become available in literature over the last decade. This includes radiative transfer models using smooth or clumpy dust and hydrodynamic models. I will review the various types of models that are currently in use and point out their main similarities, differences, and limits when it comes to interpreting observations. Finally, current and future observational challenges for these models are discussed.
%
% Do not delete next few lines
\normalsize}
\end{minipage}
%
%
%%% BODY of the paper
%%% BODY of the paper
%
\section{Introduction: Why do we need torus models? \label{sec:intro}}

The seminal work by Antonucci \& Miller \cite{Ant85} showed that the prototypical type 2 active galactic nucleus (AGN) in NGC~1068 hosts a hidden type 1 nucleus in its center. However, due to significant line-of-sight obscuration of the optical and UV emission, the broad emission lines and ``big-blue bump'' emission are not directly visible to the observer. The authors found, however, that electrons in the polar region scatter the emission from the obscured nucleus and revealed the broad-emissions typical of a type 1 AGN in the polarized flux. In this sense the electrons in the polar region act as a mirror for the broad lines.

The authors pictured the obscurer as a ``very thick absorbing disk'' without going into detail on its actual nature. It was Krolik \& Begelman \cite{Kro86} who called it the ``obscuring torus'' since a toroidal geometry captures the essence of angle-dependent, geometrically- and optically-thick obscuration. Indeed, the idea that angle-dependent obscuration is the key parameter that determines if we see an AGN as type 1 or type 2 has been unequivocally confirmed by using jet directions in radio-loud AGN as independent inclination indicators \cite{Kin00}.

It was clear early that the most likely origin of obscuration is dust. Not only does dust provide the obscuring properties, it also emits in the infrared (IR) and can be associated with the observed IR bumps in the spectral energy distributions (SEDs) of AGN \cite{Neu79}. One particular feature that strongly pointed toward dust as the origin is a generic spectral turn-over from the big-blue bump to the IR bump at 1\,$\micron$ \cite{Ede86}. Considering thermal equilibrium, this reflects well the temperature at which dust sublimates, providing a parameter-independent way of explaining this generic feature (unlike non-thermal scenarios). The same considerations also point toward a size of the order of parsecs for the obscuring dusty structure \cite{Bar87}. This brings to light a fundamental problem when investigating the torus: \textit{Although we are able to spectrally isolate the torus emission by observing in the infrared, the small sizes are beyond the spatial resolution capabilities of any single telescope.}

Finally, it has been shown that the IR power correlates well with direct luminosity indicators of the AGN, such as optical or UV broad-band luminosities or narrow and broad emission lines. More recently (e.g. see the contribution by D. Asmus) several studies showed that the AGN X-ray luminosity (supposedly originating from the inner disk corona) tightly correlates with the IR luminosity in the X-ray luminosity range of approximately $L_\mathrm{X} \sim 10^{42}-10^{45}$\,erg/s, which suggests that the ``torus'' is a universal feature with similar characteristics in all Seyfert AGN \cite{Gan09,Lev09,Asm11}.

\section{The fundamentals of torus models \label{sec:fundamentals}}

This section will briefly outline the observational and physical framework for radiative transfer models of AGN tori. The purpose of these models is usually to simulate SEDs and images to reproduce infrared observations of AGN. These models are generally static with \textit{ad hoc} assumptions of the dust distribution. A second family of models based on hydrodynamic simulations aims at self-consistently predicting the mass distribution, and the radiative transfer is only done for specific stages of the dynamic evolution. The latter models will be briefly discussed in Sect.~\ref{sec:models:hydro}.

\subsection{Observational and physical constraints}

In spite of the resolution problem, we can infer a number of constraints for the torus from indirect evidence. The basic framework of the torus -- explaining the difference between type 1 and type 2 AGN and the IR bump -- requires that the torus is \textit{dusty, obscuring (i.e. optical depth $\tau_V>1$), and geometrically thick}. Further constraints on the \textit{parsec-scaled size} can be inferred from the radiative equilibrium of dust.

Beyond these global constraints, a lot of observational and theoretical evidence suggests that the \textit{dust within the torus is arranged in clumps} instead of smoothly distributed. Indeed, the concept of a ``clumpy torus'' originates from the first theoretical considerations of the torus \cite{Kro86,Kro88}. The fact that the first models for the torus used a smooth distribution of the dust was owed to the lack of suitable computing power since radiative transfer calculations of an inhomogeneous medium are very time-expansive. 

One of the arguments for a clumpy torus considers the velocity dispersion observed in the centers of galaxies \cite{Kro88}: If these were due to the thermal motion of (smoothly-distributed) dust and gas, then the corresponding temperatures in the range of 1000s K would be too large for dust to survive. This problem can be solved easily if the velocity dispersions reflect the random motion of clouds instead. It is also argued that the high HCN/CO ratio indicates that the gas is packed into small, dense clumps, since the emitting volume of the two molecules with very different critical densities must be similar to achieve such a ratio \cite{Tac95,Tac96}. Further evidence for a clumpy torus comes from long-term variability of the column density of type 2 AGN \cite{Ris02}, although the sparse sampling provides only loose constraints on the location of the clouds responsible for these changes.

From the theoretical side, a number of first-principle arguments have been discussed that make clumpy tori plausible. As can be shown by comparing the Jeans limit with the tidal forces in the gravitational field of the supermassive black hole, small compact clouds can be stable to both self-gravity and shear forces \cite{Kro88,Bec04}. Clumpiness would also help with a more fundamental problem of the existence of the torus: The Eddington luminosity for dust around a black hole is at least 3 orders of magnitudes lower than the observed one for gas as a result of the relatively large ratio of cross section to mass (= opacity) \cite{Pie92a}. As a result dust should be efficiently removed from the nuclei of active galaxies by radiation pressure. This problem can be resolved if the dust is packed into dense clouds (with strong coupling of gas and dust) so that the radiation pressure rather acts on the cloud as a whole than on individual dust grains \cite{Hon07}.

\subsection{Principles of radiative transfer}

The physical concept behind all radiative transfer models of AGN dust tori is thermal equilibrium of dust in the radiation field of the AGN. This means that the incoming radiation is absorbed, heats up the grain, and emits thermally according to its temperature. Dust is considered a ``gray body'' for which the efficiency of absorbing, scattering, and re-emitting a photon depends on the frequency of the photon. Therefore we can write the absorbed power of a dust grain (or cell that contains dust) as
\begin{equation}
\underset {\mbox{\footnotesize incoming power}}{L_\mathrm{in}} = \int \underset {\mbox{\footnotesize absorption efficiency}}{Q_\mathrm{\nu;abs}} \times \underset {\mbox{\footnotesize incoming flux}}{F_\nu} \times \underset{\mbox{\footnotesize cross section}}{A_\mathrm{dust}} \, \mathrm{d}\nu \,.
\end{equation}
The term $Q_\mathrm{\nu;abs}F_\nu$ notes the part of the incoming flux that is actually absorbed by the dust instead of scattered or transmitted. In a similar way we can write the emitted power as
\begin{equation}
\underset {\mbox{\footnotesize outgoing power}}{L_\mathrm{out}} = \int \underset {\mbox{\footnotesize absorption efficiency}}{Q_\mathrm{\nu;abs}} \times \underset {\mbox{\footnotesize thermally-emitted flux}}{\pi B_\nu(T)} \times \underset{\mbox{\footnotesize emitting surface}}{A_\mathrm{em}} \, \mathrm{d}\nu \,.
\end{equation}
The detailed treatment and notion of these two equations depends on the exact setup of the model (e.g. single dust grain, dust cell, slab), and their solution in 1D, 2D, or 3D usually invokes a variety of different methods, from analytic solutions to iterative and non-iterative Monte-Carlo (MC) simulations and raytracing techniques.

\subsection{Model setup}

\begin{figure}
\center
\includegraphics[width=12.3cm,angle=0,clip=true]{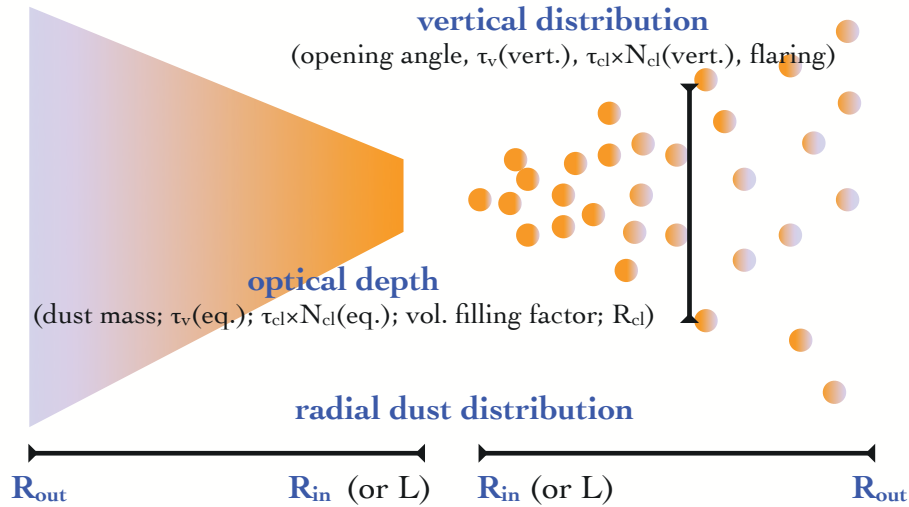} 
\caption{\label{fig:model_setup} Setup of a typical model in a clumpy (right) and smooth (left) case. Both types of models share common characteristics, with only slight adjustments to the parameters (see text for details). The sketch points out the basic model properties (blue) and outlines some versions of the actual parametrization as used in various models (black).}
\end{figure}

As mentioned previously, radiative transfer models of AGN tori use \textit{ad hoc} prescriptions for the distribution of the dust in the model space. Despite differences in the parametrization used for clumpy and smooth models, and sometimes even among the same family of models, there are fundamental similarities that capture the basics of the torus geometry and mass content in any dimensionality, as illustrated in Fig.~\ref{fig:model_setup}.

The model space is usually enclosed within an \textit{inner radius $R_\mathrm{in}$ and an outer radius $R_\mathrm{out}$}. Whether the outer radius is defined in physical units or relative to $R_\mathrm{in}$ depends on the choice of the actual model but does not make any difference. Since the dust sublimation radius (= inner radius) obeys a simple scaling law with luminosity, $R_\mathrm{in} \propto L^{1/2}$ \cite{Bar87,Sug06}, it is also possible to refer to the AGN power output instead of an absolute scaling. As a cautionary note it is important to understand that $R_\mathrm{out}$ is a model parameter that does \textit{not (!)} directly relate to an observed size. It is a mere outer boundary definition of the model space. On the contrary, choosing $R_\mathrm{out}$ too small may cause an artificial, physically unmotivated cut-off in the brightness distribution and severely influence the resulting SEDs and images (see Sect. 4.1.4 in \cite{Hon10b} for a discussion).

The \textit{distribution of the dust in radial direction} essentially defines at what radii the bulk of the dust mass is located relatively to all other radii and, therefore, sets which radii and corresponding temperatures contribute most to the overall emission (barring obscuration effects). A very common approach for the parametrization of the density distribution is a radial power law $n(r) \propto r^{-a}$ where $a$ is the power-law index that defines compactness or shallowness of the dust or dust-cloud distribution: for small $a$ ($0 \la a \la 1$), the dust mass is distributed over a wide range of distances from the AGN, while for larger $a$ the dust is concentrated toward $R_\mathrm{in}$.

Since AGN unification requires geometrical thickness, a \textit{vertical distribution of the dust} with scale height $h/r$ of the order of unity or larger has to be defined. The actual parametrization takes various shapes in literature, ranging from homogeneous vertical distributions with a cut-off height or scale-height to Gaussian or power-law distribution functions that cause a gradual fading of dust or cloud density with either height above the mid-plane or latitude. A Gaussian distribution has a physical motivation since it reflects the expected vertical density profile of an isothermal disk.

Finally, the \textit{optical depth} along any line-of-sight through the torus depends on the absolute density or dust mass in the model space. Common parameterizations use either of these values or define an optical depth value along a preferred (e.g. equatorial) line-of-site as a normalization. In case of clumpy models this optical depth is usually a combination of optical depth of an individual cloud and number of clouds along the preferred line-of-sight.

\section{An overview of AGN torus models}

Since the obscuring dust torus was established as a key component of the AGN phenomenon, a huge number of models have been presented to interpret IR observations. Over the years the models gained complexity by transitioning from smooth dust configurations, which are easy to simulate, to inhomogeneous (or clumpy) dust distributions. In this respect it should be noted again that this development is a result of increased computational power, not necessarily from new insight, because it was recognized from the beginning that the torus medium is probably clumpy. In the following, typical features and distinguishing characteristics of smooth and clumpy radiative transfer models, as well as hydrodynamical models, will be briefly discussed.

\subsection{Smooth models}

Early models for the torus used spherical, homogeneous dust distributions around the AGN or were limited to certain wavelength ranges \cite{Bar87,Row89}. Major milestones were achieved when the dust distributions became truly 2-dimension in a sense that angle-dependent obscuration was incorporated into the models and the resulting anisotropic infrared emission could be simulated \cite{Pie92b,Row93,Pie93,Efs94,Gra94,Ste94,Efs95,Man98,Fri06}, including images that can be used to predict wavelength-dependent sizes \cite{Gra97}. Although smooth models are usually 2-dimensional by design, 3D methods for the radiative transfer problem have been applied that made it easy to transition to a more inhomogeneous medium \cite{Sch05}.

Despite the fact that the smooth models are probably not correctly representing the intrinsic structure of the torus, the relative simplicity of solving the radiative transfer problem allowed for adding complexity in fields that clumpy tori do usually not support. One of these additions is temperature/distance-dependent composition of the dust. Silicate and graphite grains have very different sublimation temperatures, ranging from $800-1000$\,K for silicates and $1500-1900$\,K for graphites. In addition, at the same distance from the AGN small grains are hotter than larger grains, leading to a more complex temperature structure. Up to now, multi-species calculations have only been incorporated in smooth models \cite{Los93,Sch05}, with one exception where a clumpy model was spun-off an originally smooth-dust model \cite{Sch08}. More recently smooth models have also been used to test an alternative scenario and geometry of the dusty medium where the distribution of the dust around the AGN originates from hydrodynamic wind simulations \cite{Kea12}.

\subsection{Clumpy models \label{sec:models:clumpy}}

Radiative transfer calculations of inhomogeneous media face the problem that strong density gradients with even stronger temperature gradients are present throughout the model space. This requires very fine grid sampling of the dense regions, playing against the considerable model volume that is essentially empty. First modeling attempts were based on the dust-shell approach, as used in early smooth models, by cutting ``holes'' into the sphere \cite{Row95}. A more thorough way to approximate the average emission of a clumpy medium uses cloud source functions that are reconstructed from (2-dimensional) slab-based radiative transfer simulations, and the emerging emission through the inhomogeneous medium is calculated based on escape probabilities. More recently, advancements in computational power as well as the use of optimized methods allowed for direct Monte Carlo simulations (sometimes combined with ray-tracing techniques) of a clumpy torus, predicting both SEDs and images \cite{Dul05,Hon06,Sch08,Hon10b,Hey12}. These MC calculations also make it possible to incorporate and study the influence of inter-cloud medium, i.e. optically-thin dust that potentially occupies the void between the clouds \cite{Sta12}.

\begin{figure}
\center
\includegraphics[width=9.3cm,angle=0,clip=true]{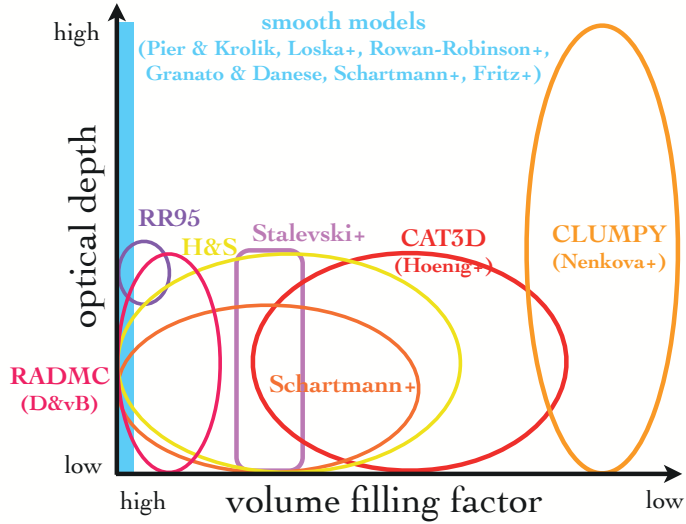} 
\caption{\label{fig:volfill} Illustration of the range of parameter space for volume filling factor and optical depth occupied by the various clumpy torus models in literature (see references in Sect.~\ref{sec:models:clumpy}). Smooth models, with unity volume filling factor, are shown in blue for reference.}
\end{figure}

Of course, there are lots of differences and assumptions in the treatment of the radiative transfer problem among the various models. To point out two distinguishing properties, Fig.~\ref{fig:volfill} compares the optical depth of the torus/dust clouds versus the volume filling factor for the clumpy models. This shows that the definition of or degree of clumpiness depends on the specific method used. MC simulations tend to have higher volume filling factors (of the order of 0.1) corresponding to larger clouds for a given distance from the AGN, while the probabilistic approach requires very low volume filling factors, i.e. very small and dense clouds. Since the combination of cloud source function (size and density) and path through the medium (global optical depth) dominates the shape of the SEDs and the appearance of the torus in an image, we may expect quite some differences between the individual models.

Another difference among the various clumpy models concerns the way the cloud density profile is treated. The emission-defining region of the cloud is the hot, directly AGN-heated surface. The bulk of the absorption and re-emission happens in this surface region leading to the highest temperature.  When looking into literature, two very different treatments of this region emerge: With increasing optical depth, this region is either transitioning toward a ``solid surface'' (steep density gradient) \cite{Nen08a,Sta12} or remains ``fluffy'' (smooth density gradient) \cite{Dul05,Hon06,Sch08}. In the former case the hot emission of optically very thick clouds becomes a black-body in the near-/mid-IR, while in the latter case the hot-surface is dominated by optically-thin emission in the infrared independent of the total density of the cloud. 

\subsection{Special purpose models}

Although not a separate family of models, a number of studies use simplified versions of a torus model to address special problems. Simplifications that are commonly utilized concern either the torus geometry (e.g. flat-disk approximation for a face-on torus) or the radiative transfer (e.g. using temperature/brightness distributions from full-scale models). The propagation of AGN variability through the dust medium has been a successful application of these kind of models, showing that the distribution is an important parameter that shapes the near-IR and mid-IR light curves \cite{Bar92,Kaw10,Hon11}. Another point-of-interest has been the effect of anisotropic accretion disk radiation on the IR emission and dust distribution at the inner rim of the torus \cite{Kaw11}.

\subsection{Hydrodynamic models \label{sec:models:hydro}}

The advantage of radiative transfer models is their relative simplicity that makes it easy to simulate model grids. These grids can then be used to reproduce observations. However, they do not contain any physics. In order to study the dynamics of the gas and dust around the black hole in the region of the torus, a number of hydrodynamic models have been developed with the goal to reproduce the mass distribution self-consistently \cite{Wad02,Wad05,Dor08a,Dor08b,Sch09,Wad09,Dor09,Per11,Hop12a,Hop12b}. This usually comes at the expense of the radiative transfer, making it difficult to model photometric observations for a wide range of parameters. On the other hand, since both densities and kinematics are predicted, the models provide a physical basis for studying molecular lines on scales of several to 10s of parsecs.

Although the details vary among the different models, it has been shown that the gas in the inner 5-10\,pc around the AGN collapses to a thin disk \cite{Sch09,Wad09}. Outside of this inner dynamically cool disk the turbulence in the clumpy gas can provide for the necessary geometrically-thick obscuration. However, the exact vertical distribution and substructure does depend on the boundary conditions of the simulations.

It has been realized recently that radiation pressure of the AGN on the circumnuclear medium, but also of the thermal radiation from the gas and dust itself, can have a huge effect on the distribution of the material in the inner torus region and the stability of dust clumps \cite{Kro07,Sch11}. Consequently it is now attempted to work radiation pressure into the hydrodynamical solution \cite{Dor11,Dor12,Wad12}. One of the big problems is that in order to properly account for both external and internal radiation pressure, the radiative transfer through the medium has to be solved in some way, which adds huge overheads on the already time-consuming calculations.

\section{Caveats when using torus models}

Given the number of radiative transfer models that are available, they have become popular to model infrared observations of AGN. Although this is a very desirable task and potentially adds value to the interpretation, it is very important to keep in mind some very fundamental restrictions of the models:

\begin{itemize}
\item \textit{Modeling means observing the model.} As previously mentioned, there are different ways to setup a model and treat the radiative transfer that affect the SED. As such, if a best-fitting model is found for a set of photometric observations, these models are only relevant in the context of the model and do not (necessarily) tell us something about the real object. This fundamental problem originates from the fact that the infrared emission is essentially a weighted combination of gray-body emission. The weights are determined by the distribution of the dust and the intrinsic obscuration, and the same SED can be obtained by different configurations/combinations. Indeed, it is well possible that we will find a very good fit of the SED from a model that does not obey the basic requirements for an AGN torus (see Sect.~\ref{sec:fundamentals}), in particular if the observations lack spatial information.

\item \textit{Different models can come to contradictory conclusions.} The previous problem may also manifest in the answer we may get from two different torus models of the same family (i.e. smooth or clumpy). Given the various treatments of the radiative transfer problem and other ``hidden'' aspects (e.g. as described in the last paragraph of Sect.~\ref{sec:models:clumpy}), it is not surprising that the same set of parameters can lead to very different SEDs. One example is the appearance of the silicate feature in the different clumpy models presented in \cite{Nen08b} (Fig. 15, bottom-left panel) and \cite{Hon10b} (Fig. 7, top-right panel). Even when all other parameters are kept approximately the same, face-on views onto the torus results in an absorption feature in the former and an emission feature in the latter model (see the talk slides for a one-on-one comparison). Moreover, it was shown that two very different smooth and clumpy models can model the same set of observations equally well, however never for the same set of parameters \cite{Fel12}. As such, conclusions drawn from data fitting or parameter studies within the a given model should be considered model specific, even within the same family of models.

\item \textit{Model parameters are often degenerate.} Finding the best fit to the model does not necessarily mean that the parameters are unique. In fact, since observational data are limited and the number of model parameters significant, it can be expected that similarly good-fitting solution can be found. One such combination of models is the opening angle, optical depth, and inclination that suffers from the fact that our view of an AGN is limited to one specific line-of-sight. A recent study that models photometric and spectroscopic IR data exemplifies the degree of parameter degeneracies that can be expected by analyzing 2-dimensional probability distribution functions of all pairwise combinations of parameters in one specific model \cite{Alo11}.
\end{itemize}

\section{Future challenges and possibilities}

Torus models that are currently in use face a lot of challenges from different sides. First, as Andy Lawrence felicitously summarized: ``We have a lot of models, can we have a theory for them now?'' Indeed, the connection of the \textit{ad hoc} parameterizations of the radiative transfer models to more fundamental physics or results from (radiation-)hydrodynamic models is poorly explored. It is very likely that some of the configurations currently used are dynamically unstable (see contribution by M. Schartmann). Moreover, simulations and analytic models have shown that dust clumps exposed to the direct radiation of the AGN dissolve over time or are rather expelled than accreted \cite{Hon07,Sch11}. Although, admittedly, the hydrodynamic models are still being worked on to include all relevant physical processes, it may be beneficial to keep an eye on simulation results when setting up radiative transfer models, in particular in the realm that is not easily accessible or constrained by observations (e.g. dust cloud properties).

The more direct challenges for current torus models arise from observations. With the advent of both infrared interferometry and high-quality photometric and spectroscopic surveys, our knowledge of the infrared emission of AGN has exploded. The Spitzer satellite provided mid- to far-IR  data for a huge number of AGN of all classes. One of the features that emerged from these observations was a distinct near-IR bump in type 1 AGN, peaking about $3-5\,\micron$; in fact, this near-IR bump has already been seen in earlier datasets but much less pronounced \cite{Ede86,Mor09,Mor12}. Quantitatively reproducing this bump using torus models has proven to be a challenge and it is generally modelled by incorporating a separate hot graphite-dust component. Although it has been suggested that this bump is caused by the fact that the dust sublimation radius is rather a sublimation region of different grain sizes and species, models that do account for multi-grain radiative transfer did not produce a significant bump \cite{Sch05,Sch08}. It seems that there is great potential in improving the models and enhance our understanding of the geometry and composition of the hot dust by trying to reproduce this spectral feature.

The biggest step forward for improving models can probably be made by taking into account the spatial information from the growing number of IR-interferometrically-observed and -resolved AGN. For that, models have to calculate images and simulate the interferometric observations based on these images in Fourier space; relying on ``size'' comparisons may not be sufficient or even misleading \cite{Kis11b}. Interferometry mainly provides information about the brightness distribution of the torus emission projected onto the plane of the sky. With the next generation of interferometric instruments (MATISSE, GRAVITY), additional such data will become available. Knowing the wavelength-dependent brightness distribution and comparing it to predictions by the model helps to eliminate degeneracies from flux-only data. Efforts to use radiative transfer models in this way have been undertaken only in the cases of NGC~1068, the Circinus galaxy, and NGC~3783 up to now \cite{Hon06,Sch08,Hon10a}.

%
%
% Do not delete the next line
\small  % Do not delete
%
%%% Comment the following line if you do not have acknowledgments.
\section*{Acknowledgments}   % Do not delete if you declare acknowledgments
%
%%% ACKNOWLEDGMENTS
%%% ACKNOWLEDGMENTS
The author acknowledges support by Deutsche Forschungsgemeinschaft (DFG) in the framework of a research fellowship (Auslandsstipendium).

%
% Do not delete the next few lines

%

\begin{thebibliography}{}
\small
%
%%% BIBLIOGRAPHY
%%% BIBLIOGRAPHY
\bibitem{Alo11}{Alonso-Herrero, A., et al. 2011, 736, 82}
\bibitem{Ant85}{Antonucci, R.~R.~J., \& Miller, J.~S. 1985, ApJ, 297, 621}
\bibitem{Asm11}{Asmus, D., et al. 2011, A\&A, 536, 36}
\bibitem{Bar87}{Barvainis, R. 1987, ApJ, 320, 537}
\bibitem{Bar92}{Barvainis, R. 1992, ApJ, 400, 502}
\bibitem{Bec04}{Beckert, T., \& Duschl, W.~J. 2004, A\&A, 426, 445}
\bibitem{Dor08a}{Dorodnitsyn, A., et al. 2008a, ApJ, 675, L5}
\bibitem{Dor08b}{Dorodnitsyn, A., et al. 2008b, ApJ, 687, 97}
\bibitem{Dor09}{Dorodnitsyn, A. \& Kallman, T. 2009, ApJ, 703, 1797}
\bibitem{Dor11}{Dorodnitsyn, A., et al. 2011, ApJ, 741, 29}
\bibitem{Dor12}{Dorodnitsyn, A., et al. 2012, ApJ, 747, 8}
\bibitem{Dul05}{Dullemond, C.~P., \& van Bemmel, I. 2005, A\&A, 436, 47}
\bibitem{Ede86}{Edelson, R.~A., \& Malkan, M.~A. 1986, ApJ, 308, 59}
\bibitem{Efs94}{Efstathiou, A. \& Rowan-Robinson, M. 1994, MNRAS, 266, 212}
\bibitem{Efs95}{Efstathiou, A. \& Rowan-Robinson, M. 1995, MNRAS, 273, 649}
\bibitem{Fel12}{Feltre, A., et al. 2012, MNRAS, 426, 120}
\bibitem{Fri06}{Fritz, J., et al. 2006, MNRAS, 366, 767}
\bibitem{Gan09}{Gandhi, P., et al. 2009, A\&A, 502, 457}
\bibitem{Gra94}{Granato, G.~L. \& Danese, L. 1994, MNRAS, 268, 235}
\bibitem{Gra97}{Granato, G.~L., et al. 1997, ApJ, 486, 147}
\bibitem{Hey12}{Heymann, F., \& Siebenmorgen, R. 2012, ApJ, 751, 27}
\bibitem{Hon06}{H\"onig, S.~F., et al. 2006, A\&A, 452, 459}
\bibitem{Hon07}{H\"onig, S.~F., \& Beckert, T. 2007, MNRAS, 380, 1172}
\bibitem{Hon10a}{H\"onig, S.~F., et al. 2010, A\&A, 515, 23}
\bibitem{Hon10b}{H\"onig, S.~F., \& Kishimoto, M. 2010, A\&A, 523, 27}
\bibitem{Hon11}{H\"onig, S.~F. \& Kishimoto, M. 2011, A\&A, 534, 121}
\bibitem{Hop12a}{Hopkins, P.~F., et al. 2012a, MNRAS, 420, 320}
\bibitem{Hop12b}{Hopkins, P.~F., et al. 2012b, MNRAS, 425, 1121}
\bibitem{Kaw10}{Kawaguchi, T. \& Mori, M. 2010, ApJ, 724, L183}
\bibitem{Kaw11}{Kawaguchi, T. \& Mori, M. 2011, ApJ, 737, 105}
\bibitem{Kea12}{Keating, S., et al. 2012, ApJ, 749, 32}
\bibitem{Kin00}{Kinney, A.~L., et al. 2000, ApJ, 537, 152}
\bibitem{Kis11b}{Kishimoto, M., et al. 2011, A\&A, 536, 78}
\bibitem{Kro86}{Krolik, J.~H., \& Begelman, M. C. 1986, ApJ, 308, L55}
\bibitem{Kro88}{Krolik, J.~H., \& Begelman, M. C. 1988, ApJ, 329, 702}
\bibitem{Kro07}{Krolik, J.~H. 2007, ApJ, 661, 52}
\bibitem{Lev09}{Levenson, N.~A., et al. 2009, ApJ, 703, 390}
\bibitem{Los93}{Loska, Z., et al. 1993, MNRAS, 261, 63}
\bibitem{Man98}{Manske, V., et al. 1998, A\&A, 331, 52}
\bibitem{Mor09}{Mor, R., et al. 2009, ApJ, 705, 298}
\bibitem{Mor12}{Mor, R., \& Netzer, H. 2012, MNRAS, 420, 526}
\bibitem{Nen02}{Nenkova, M., et al. 2002, ApJ, 570, L9}
\bibitem{Nen08a}{Nenkova, M., et al. 2008a, ApJ, 685,147}
\bibitem{Nen08b}{Nenkova, M., et al. 2008b, ApJ, 685, 160}
\bibitem{Neu79}{Neugebauer, G., et al. 1979, ApJ, 230, 79}
\bibitem{Per11}{Perez-Beaupuits, J., et al. 2011, ApJ, 730, 48}
\bibitem{Pie92a}{Pier, E.~A. \& Krolik, J.~H. 1992a, ApJ, 401, 99}
\bibitem{Pie92b}{Pier, E.~A. \& Krolik, J.~H. 1992b, ApJ, 401, 99}
\bibitem{Pie93}{Pier, E.~A. \& Krolik, J.~H. 1993, ApJ, 418, 673}
\bibitem{Ris02}{Risaliti, G., et al. 2002, ApJ, 571, 234}
\bibitem{Row89}{Rowan-Robinson, M., \& Crawford, J. 1989, MNRAS, 238, 523}
\bibitem{Row93}{Rowan-Robinson, M., \& Efstathiou, A. 1993, MNRAS, 263, 675}
\bibitem{Row95}{Rowan-Robinson, M. 1995, MNRAS, 272, 737}
\bibitem{Sch05}{Schartmann, M., et al. 2005, A\&A, 437, 861}
\bibitem{Sch08}{Schartmann, M., et al. 2008, A\&A, 482, 67}
\bibitem{Sch09}{Schartmann, M., et al. 2009, MNRAS, 393, 759}
\bibitem{Sch11}{Schartmann, M., et al. 2011, MNRAS, 415, 741}
\bibitem{Sta12}{Stalevski, M., et al. 2012, MNRAS, 420, 2756}
\bibitem{Ste94}{Stenholm, L. 1994, A\&A, 290, 393}
\bibitem{Sug06}{Suganuma, M., et al. 2006, ApJ, 639, 46}
\bibitem{Tac95}{Tacconi, L.~J., et al. 1995, ApJ, 426, L77}
\bibitem{Tac96}{Tacconi, L.~J. 1996, IAUS, 178, 489}
\bibitem{Wad02}{Wada, K., \& Norman, C.~A. 2002, ApJ, 566, L21}
\bibitem{Wad05}{Wada, K., \& Tomisaka, K. 2005, ApJ, 619, 93}
\bibitem{Wad09}{Wada, K., et al. 2009, ApJ, 702, 63}
\bibitem{Wad12}{Wada, K., 2012, ApJ, 758, 66}
%
%
% Do not delete next few lines
\end{thebibliography}
\end{document}